\documentclass{webofc}
\usepackage[varg]{txfonts}   
\usepackage{graphicx}
\usepackage{geometry}
\usepackage{fancyhdr}
\usepackage{epsfig}
\usepackage[compat=1.1.0]{tikz-feynman}
\begin{document}
%
%
\title{SIMP dark matter and its cosmic abundances}
%
\author{\firstname{Soo-Min} \lastname{Choi}\inst{1}\fnsep\thanks{\email{soominchoi90@gmail.com}} \and
        \firstname{Hyun Min} \lastname{Lee}\inst{1}\fnsep\thanks{\email{hminlee@cau.ac.kr}} \and
        \firstname{Min-Seok} \lastname{Seo}\inst{2}\fnsep\thanks{\email{minseokseo57@gmail.com}}
        }
%
\institute{Department of Physics, Chung-Ang University, Seoul 06974, Korea. 
\and
           Center for Theoretical Physics of the Universe, Institute for Basic Science, 34051 Daejeon, Korea. 
          }
%
\abstract{
We give a review on the thermal average of the annihilation cross-sections for $3\rightarrow 2$ and general higher-order processes. Thermal average of higher order annihilations highly depend on the velocity of dark matter, especially, for the case with resonance poles. We show such examples for scalar dark matter in gauged $Z_3$ models.
}
\maketitle
\vspace{0.3cm}
\begin{center}
{\it \small Prepared for the proceedings of the 13th International Conference on Gravitation,\\
Ewha Womans University, Korea, 3-7 July 2017.}
\end{center}
\vspace{0.3cm}
%
%
\section{Introduction}
%
\label{intro}
Dark matter has been hinted by missing masses in galaxies and galaxy clusters and supported by indirect evidences such as Cosmic Microwave Background(CMB) or gravitational lensing effect. So many experiments have been carried out to search the dark matter, through direct detection, indirect detection and collider search. Although we do not know characteristics of dark matter yet, we know the non-gravitational interaction of dark matter is getting more constrained.
\par %
The most popular candidate for dark matter is Weakly Interacting Massive Particle(WIMP), and WIMP has $2 \rightarrow 2$ annihilation channels into the Standard Model particles. WIMP relic density is determined when the $2 \rightarrow 2$ annihilation channels are frozen out. There have been a lot of experiments undertaken to find the WIMP, but we have not found any experimental evidence to date.
\par %
However, to solve the small scale problems of collisionless cold dark matter, people consider the self-interaction between dark matters as one of the solutions. Small scale problems stem from the differences between observation and simulation of the dark matter profiles, the density of sub-halos, and the number of sub-halos. The popular candidate for self-interacting dark matter is Strongly Interacting Massive Particles(SIMP) \cite{Hochberg1}. SIMP has a $ 3 \rightarrow 2 $ self-annihilation process between dark matter particles determining the current relic density when it is frozen-out. Unlike the WIMP, SIMP doesn't need to have a large coupling with the Standard Model particles. The mass of SIMP is order of 100 MeV scale so it is much lighter than WIMP.
\par %
In order to accurately determine the relic density of dark matter, a thermal averaged cross-section of the annihilation channel is required. For the non-relativistic dark matter, the cross-section can be expanded as powers of dark matter relative velocity and the lowest order terms of relative velocity are the most important. This applies equally to high order annihilations such as SIMP, especially when there is a higher partial wave or resonance pole.
\par %
In this article, we discuss the thermal averaged cross-section of high order annihilations for SIMP and perform a valid thermal average of the velocity dependent cross-section such as in the case with resonance pole.
%
%
\section{Resonance and thermal average}
%
\label{sec-2}
In order to obtain the relic density of dark matter, we need to calculate a thermal average of cross-section with the integration of a momenta for initial particles. For the non-relativistic dark matter, if the cross-section has a resonance of Breit-Wigner form and the sum of masses for initial particles are almost the same as the resonance mass, the thermal averaged cross-section is very sensitive to the velocities of initial dark matter particles. So we need to consider exactly the momentum dependence of the cross-section. In the paper of Gondolo and Gelmini \cite{gondolo}, they showed the exact results of thermal average only for $2\rightarrow 2$ cross-section.
\par %
Recently, SIMP and higher order annihilating models have shed light on solving the small scale problems \cite{simpe1,simpe2,simpe3,simpe4,simpe5,simpe6,simpe7}. To get the right relic density, the exact calculation of thermal average for $n\rightarrow 2$ $(n\geq 3)$ cross-section is also very important. So in this article, we show the results of more general thermal average of both non-resonant and resonant cross-sections.
%
%
\section{Thermal average of SIMP}
%
\label{sec-3}
An interesting way for producing the self-interacting dark matter is through higher order annihilations, namely, the initial particles are three or more in annihilation channels. A good example of self-interacting dark matter with higher order annihilation is SIMP. The relic density of SIMP is governed by the Boltzmann equation,
\begin{equation}
\begin{split}
\dot{n}_\chi+3Hn_\chi=-\langle \sigma v^2 \rangle (n_\chi^3-n_\chi n_{eq}^2)
\end{split}
\end{equation}
where
\begin{equation}
\begin{split}
\sigma v^2=\frac{1}{2E_1 2E_2 2E_3}\int \frac{d^4p_4}{(2\pi)^3 2E_4}\frac{d^4p_5}{(2\pi)^3 2E_5}|\mathcal{M}|^2(2\pi)^4\delta^4(p_1+p_2+p_3-p_4-p_5).
\end{split}
\end{equation}
\par %
The cross-section with 5-point interactions should be thermal averaged to get the exact relic density. Thermal averaged $3\rightarrow 2$ cross-section is very similar with $2\rightarrow 2$ cross-section. We just need to consider one more initial dark matter. For the initial particles with the same masses, the thermal averaged $3\rightarrow 2$ cross-section is given by \cite{simpe5},
\begin{equation}
\langle \sigma v^2\rangle= \frac{\int d^3 v_1 d^3 v_2 d^3 v_3\, \delta^3({\vec v}_1+{\vec v}_2+{\vec v}_3)(\sigma v^2)\, e^{-\frac{1}{2}x(v^2_1+v^2_2+v^2_3)}}{\int d^3 v_1 d^3 v_2 d^3 v_3\, \delta^3({\vec v}_1+{\vec v}_2+{\vec v}_3) \,e^{-\frac{1}{2}x(v^2_1+v^2_2+v^2_3)}}.
\end{equation}
where delta function comes from center of mass frame of initial particles. If the 5-point interaction is non-resonant, then the cross-section can be expanded by the velocities of initial particles like,
\begin{equation}
(\sigma v^2)= a_0 + a_1 (v^2_1+v^2_2+v^2_3) + a^{(1)}_2 (v^2_1+v^2_2+v^2_3)^2 + a^{(2)}_2 (v^4_1+v^4_2+v^4_3)+\cdots .
\end{equation}
\par %
So there are SO(9) invariant terms and general terms of the form, $(v^2_1)^n(v^2_2)^m(v^2_3)^l$. If the $3\rightarrow 2$ cross-section has only SO(9) invariant terms, the corresponding thermal average is
\begin{equation}
\begin{split}
\langle \sigma v^2\rangle&= \frac{1}{2} x^3 \sum_{l=0}^\infty \frac{a_l}{l!}\,\int^\infty_0 d\eta\, \eta^{l+2} e^{-x\eta} \\
&= a_0+3 a_1 x ^{-1}+6 a_2 x^{-2}+\cdots . \label{etan}
\end{split}
\end{equation}
with $\eta\equiv \frac{1}{2}(v^2_1+v^2_2+v^2_3)$. Otherwise, if the cross-section depends on the general terms, then,
\begin{equation*}
\begin{split}
\langle(v^2_1)^n(v^2_2)^m(v^2_3)^l\rangle &= \frac{3\sqrt{3} x^3}{\pi}\, \int^\infty_0 dv_1 v^2_1 \int^\infty_0 dv_2 v_2^2 (v^2_1)^n (v^2_2)^m \times \\
& \quad \times \int^{+1}_{-1} d\cos\theta_{12} (v^2_1+v^2_2+2v_1 v_2 \cos\theta_{12})^l \, e^{-x(v^2_1+v^2_2+v_1 v_2 \cos\theta_{12})},
\end{split}
\end{equation*} 
especially,
\begin{equation*}
\langle(v^2_1)^n\rangle=\Bigg(\frac{4}{3}\Bigg)^n \frac{\Gamma(n+\frac{3}{2})}{\Gamma(\frac{3}{2})}\, x^{-n}= \langle(v^2_2)^n\rangle= \langle(v^2_3)^n\rangle. \label{v1n}
\end{equation*}
\par %
When the $3 \rightarrow 2$ process is due to resonance, then the resulting cross-section has also a Breit-Wigner form like,
\begin{equation}
\begin{split}
\langle \sigma v^2 \rangle &= \Bigg\langle \sum_{l=0}^\infty \frac{b^{(l)}_R}{l!}\,\eta^l \,\frac{\gamma_R}{(\epsilon_R-\frac{2}{3}\eta)^2+\gamma^2_R} \Bigg\rangle\\
&=\frac{3}{4}\pi x^3\sum_{l=0}^\infty \frac{b^{(l)}_R}{l!}\, G_l(z_R;x), 
\end{split}
\end{equation}
where
\begin{equation}
\begin{split}
G_l(z_R;x)&= {\rm Re}\bigg[\frac{i}{\pi}\int^\infty_0 d\eta\, \frac{ \eta^{l+2} e^{-x\eta}}{\frac{3}{2}z_R-\eta} \bigg] \\
&= (-1)^l \frac{\partial^l}{\partial x^l}\, G_0(z_R;x),
\end{split}
\end{equation}
$\epsilon_R\equiv\frac{m^2_R-9m^2_{\rm DM}}{9m^2_{\rm DM}}$, $\gamma_R\equiv \frac{m_R \Gamma_R}{9m^2_{\rm DM}}$ and $z_R\equiv \epsilon_R +i\gamma_R$. When $\gamma_R\ll 1$, $G_0(z_R; x)\approx \frac{9}{4} \epsilon^2_R e^{-\frac{3}{2} x \epsilon_R}\theta(\epsilon_R)$. So the thermal averaged cross-section is given by
\begin{equation}
\langle\sigma v^2 \rangle_R\approx \frac{27}{16}\pi\epsilon^2_R  x^3e^{-\frac{3}{2} x \epsilon_R}\theta(\epsilon_R)  \sum_{l=0}^\infty \frac{b^{(l)}_R}{l!}\, \Big(\frac{3}{2}\Big)^l \epsilon^l_R. \label{nwa}
\end{equation}
Because the phase space integral in the thermal average for SIMP depends on the higher order of dark matter velocity than in the case of WIMP, the SIMP case highly depends on the resonance mass($\epsilon^2_R$) than WIMP case($\epsilon^{1/2}_R$).
\par %
We can generalize the above results of the thermal average of SIMP processes for non-degenerate initial particle masses and even higher order annihilation like $n\rightarrow 2,$ where $n>3$. Even if the masses of the initial particles are different, we can calculate the thermal averaged cross-section. Especially, when the cross-section is s-wave, the result can be obtained by $m\rightarrow (m_1+m_2+m_3)/3$ in the definition of x in eq.(12), (13) and (14). For even higher order annihilations, we can get the general p-wave results by using SO(3n) symmetry.
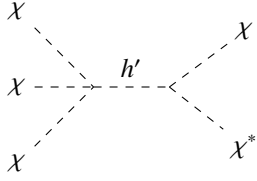
\begin{figure}[h]
\centering
\begin{center}
\begin{tikzpicture}
\begin{feynman}
    \vertex (i1) {\(\chi\)};
    \vertex[below = 1cm of i1] (i2) {\(\chi\)};
    \vertex[below = 1cm of i2] (i3) {\(\chi\)};
    \vertex[right = 1cm of i2] (l1);
    \vertex[right = 1cm of l1] (l2);
    \vertex[right = 1cm of l2] (l3);
    \vertex[above = 0.5cm of l3] (f1) {\(\chi\)};
    \vertex[below = 0.5cm of l3] (f2) {\(\chi^\ast\)};
    \diagram* {
      (i1) -- [scalar] (l1) -- [scalar, edge label=\(h'\)] (l2) -- [scalar] (f1),
      (i2) -- [scalar] (l1),
      (i3) -- [scalar] (l1),
      (l2) -- [scalar] (f2),
    };
\end{feynman}
\end{tikzpicture}
\end{center}
\caption{One example of the dark Higgs resonance channel of the discrete $Z_3$ gauged symmetry dark matter model}
\label{fig1}
\end{figure}
%
%
\section{Model for resonant SIMP}
\begin{figure}
  \begin{center}
   \includegraphics[height=.4\textwidth]{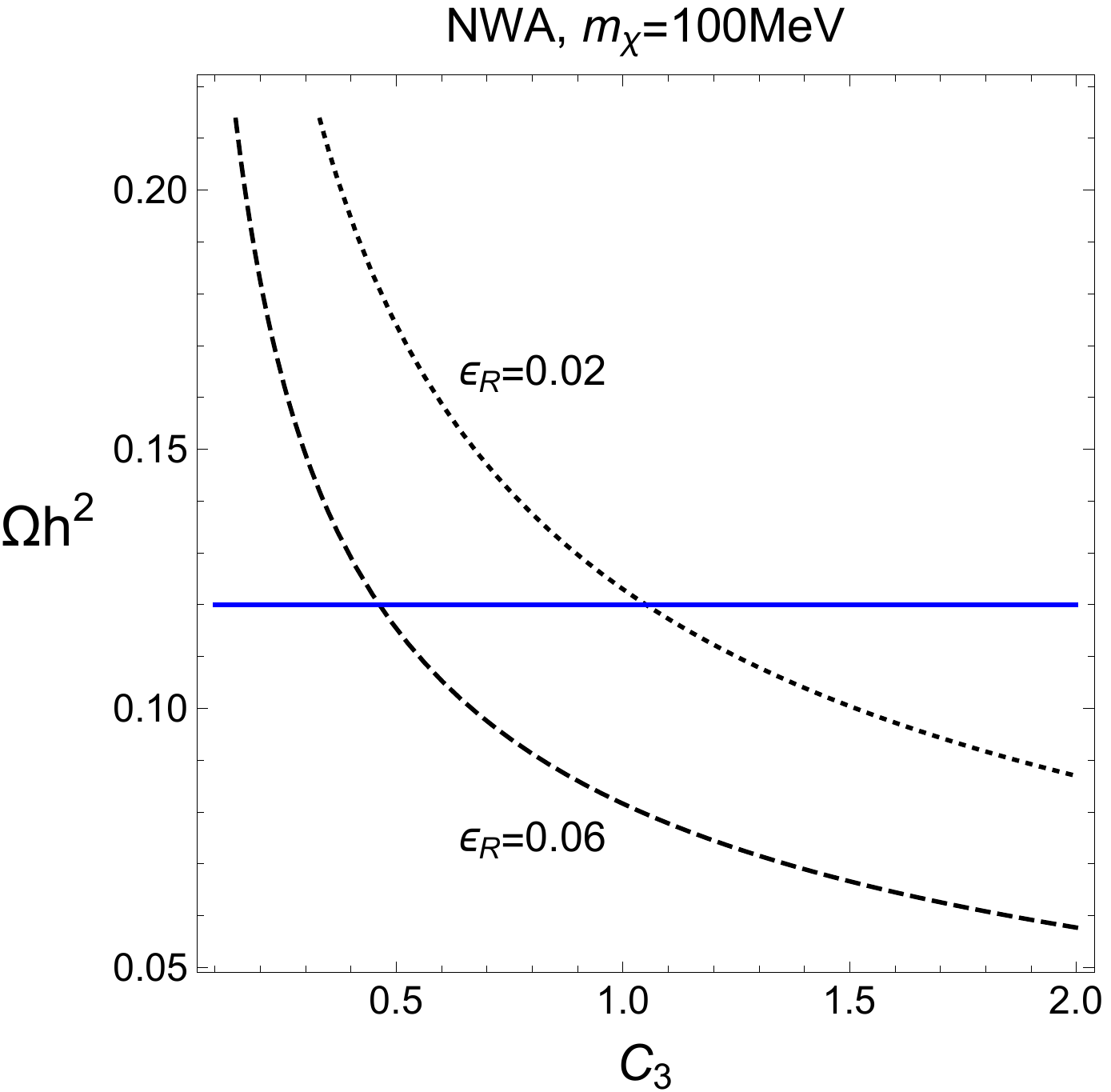}
    \quad\quad \includegraphics[height=.4\textwidth]{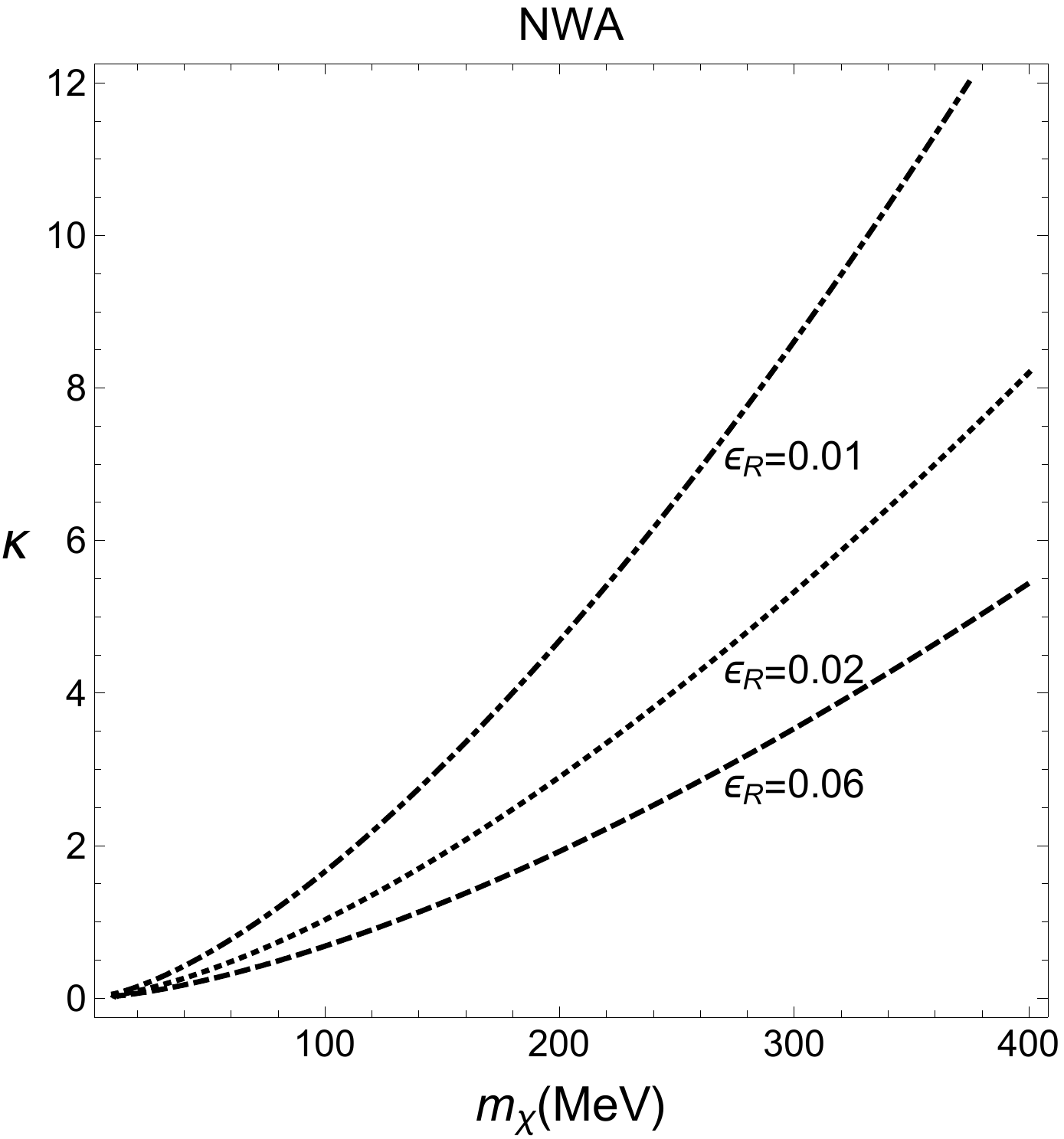}
   \end{center}
  \caption{Relic density of $Z_3$ dark matter as a function of $C_3$ and $\epsilon_R$ in the left. Dark matter cubic coupling satisfying a correct relic density as a function of dark matter mass and $\epsilon_R$ in the right. The blue line is the relic density of dark matter obtained by Planck.}
  \label{fig2}
\end{figure}

%
\label{sec-4}
As a concrete model, we show the discrete $Z_3$ gauged model for scalar dark matter. In this model, there are dark Higgs and dark matter which are charged under the dark U(1) gauge symmetry. A dark matter triple coupling $\kappa$ is generated after the dark U(1) is broken down to $Z_3$. The details of the model are in the references \cite{simpe3,simpe6}. Because there are dark Higgs resonance channels as shown in figure~\ref{fig1}, we should use the results near resonance. In the non-relativistic limit, the effective $3\rightarrow 2$ annihilation cross-section with a dark Higgs resonance has the form,
\begin{equation}
\begin{split}
(\sigma v^2)_{Z_3}= C_3\frac{\gamma_R}{(\epsilon_R-\frac{2}{3}\eta)^2+\gamma_R^2},
\end{split}
\end{equation}
and $C_3$ is given by
\begin{equation}
\begin{split}
C_3=\frac{\sqrt{5}\kappa^2}{12\beta_\chi m_\chi^5}\Bigg(1+\frac{\lambda_{\phi\chi}v'^2}{m_\chi^2}\Bigg)^2.
\end{split}
\end{equation}
Definition of $\beta_{\chi}$ is $\sqrt{1-4m_\chi^2/m_{h'}^2}$. Because the three-body decay ($h'\rightarrow \chi \chi \chi$) is suppressed by extra phase space with $\epsilon_R^2/(4\pi^2)$, the decay width of dark Higgs is determined mainly by two-body decay ($h'\rightarrow \chi \chi^\ast$). From eq.(6), the thermal averaged cross-section of eq.(9) is given by
\begin{equation}
\begin{split}
\langle\sigma v^2 \rangle_{Z_3}=\frac{3}{4}C_3\pi x^3G_0(z_R;x).
\end{split}
\end{equation}
By using the narrow width approximation,
\begin{equation}
\begin{split}
\langle\sigma v^2 \rangle_{Z_3} \approx \frac{27}{16}C_3\pi \epsilon_R^2x^3e^{-\frac{3}{2}x\epsilon_R}\theta(\epsilon_R).
\end{split}
\end{equation}
In figure~\ref{fig2}, we show the parameter space for the relic density of dark matter as a function of $C_3$ and $\epsilon_R$ in the left panel. We choose dark matter mass as 100 MeV. Right panel is the parameter space for dark matter cubic coupling $\kappa$ and dark matter mass satisfying the relic density of dark matter. In this plot, we show the value of $\epsilon_R=0.01, 0.02,0.06$. For both plots in figure~\ref{fig2}, we assume narrow width approximation(NWA).
%
%
\section{Conclusions}
%
\label{sec-5}
We have shown the general results of the thermal averaged cross-sections for SIMP and higher order annihilations. When the cross-section has high dependence on velocity and resonance pole, the exact calculation of thermal averaged cross-section is very important for the relic density of dark matter. Moreover, the results can be generalized for the non-degenerate initial particle masses and even higher annihilation channels.
%
%
\section{Acknowledgments}
%
\label{sec-6}
The work is supported in part by Basic Science Research Program through the National Research Foundation of Korea (NRF) funded by the Ministry of Education, Science and Technology (NRF-2016R1A2B4008759). The work of SMC is supported in part by TJ Park Science Fellowship of POSCO TJ Park Foundation. MS is supported by IBS under the project code, IBS-R018-D1.
%
%

%

\begin{thebibliography}{}
%
%
\bibitem{Hochberg1}
  Y.~Hochberg, E.~Kuflik, T.~Volansky and J.~G.~Wacker,
  Phys.\ Rev.\ Lett.\  \textbf{113} 171301 (2014)
%
\bibitem{gondolo}
  P.~Gondolo and G.~Gelmini,
  Nucl.\ Phys.\ B {\bf 360} 145. (1991) 
%
\bibitem{griest}
  K.~Griest and D.~Seckel,
  Phys.\ Rev.\ D {\bf 43} 3191. (1991)
%
 \bibitem{simpe1} 
  Y.~Hochberg, E.~Kuflik, H.~Murayama, T.~Volansky and J.~G.~Wacker,
  Phys.\ Rev.\ Lett.\  {\bf 115} 2, 021301 (2015) 
%
\bibitem{simpe2}
  N.~Bernal, C.~Garcia-Cely and R.~Rosenfeld,
  JCAP \textbf{1504} 04,  012 (2015)
%
\bibitem{simpe7}
  H.~M.~Lee and M.~S.~Seo,
  Phys.\ Lett.\ B \textbf{748} 316 (2015)
%
\bibitem{simpe3}
  S.~M.~Choi and H.~M.~Lee,
  JHEP {\bf 1509} 063 (2015)
%
\bibitem{simpe4}
  S.~M.~Choi and H.~M.~Lee,
  Phys.\ Lett.\ B {\bf 758} 47 (2016)
%
\bibitem{simpe5}
  S.~M.~Choi, H.~M.~Lee and M.~S.~Seo,
  JHEP {\bf 154} 04 (2017)
%
\bibitem{simpe6}
  S.~M.~Choi, Y.~J.~Kang and H.~M.~Lee,
  JHEP {\bf 1612} 099 (2016)
%
\end{thebibliography}
\end{document}